\documentstyle[12pt]{article}
\input epsf
\newcommand{\beq}{\begin{eqnarray}}
\newcommand{\eeq}{\end{eqnarray}}

\newcommand{\gsim}{\lower.7ex\hbox{$
\;\stackrel{\textstyle>}{\sim}\;$}}
\newcommand{\lsim}{\lower.7ex\hbox{$
\;\stackrel{\textstyle<}{\sim}\;$}}

\input epsf
\newcommand{\grpicture}[1]
{
    \begin{center}
        \epsfxsize=300pt
        \epsfysize=200pt
        \vspace{-5mm}
        \parbox{\epsfxsize}{\epsffile{#1.eps}}
        \vspace{5mm}
    \end{center}
}

\begin{document}
\begin{titlepage}

\begin{flushright}
ITEP-TH-3/98 \\

\end{flushright}
\vspace*{3cm}

\begin{center}
{\Large \bf   BPS and non--BPS Domain Walls   in Supersymmetric $QCD$
with $SU(3)$ Gauge Group}
\vspace{2cm}

{\Large  A.V. Smilga and A.I. Veselov} \\

\vspace{0.8cm}

{\it ITEP, B. Cheremushkinskaya 25, Moscow 117218, Russia}\\

\end{center}

\vspace*{2cm}

\begin{abstract}
We study the spectrum of the domain walls
interpolating between different chirally asymmetric vacua
in supersymmetric QCD with the $SU(3)$ gauge group and
including 2 pairs of chiral matter multiplets in fundamental
and anti-fundamental representations.
 For small enough masses $m < m_* \approx .286 \ \Lambda_{SQCD}$, there
 are two different domain wall solutions which are BPS--saturated and two
 types of ``wallsome sphalerons''.
  At $m = m_*$, two BPS branches join together and, in the interval
 $m_* < m < m_{**} \approx 3.704 \ \Lambda_{\rm SQCD}$,
 BPS equations have no solutions but there are solutions to the
  equations of motion
describing a non--BPS domain wall and a sphaleron. For $m > m_{**}$,
there are no solutions whatsoever.
\end{abstract}

\end{titlepage}

\section{Introduction}

Supersymmetric QCD is the theory involving  a gauge
 vector supermultiplet
$V$ and some number of chiral matter supermultiplets.
The models of this class attracted  attention of theorists since the
beginning of the
eighties and many interesting and non--trivial results concerning their
non--perturbative dynamics  have been obtained \cite{brmog}.
The dynamics depends in an essential way on the gauge group, the matter
content, the masses of the matter fields and their Yukawa couplings.

 The most simple  in some sense variant of the
model is based on the $SU(N)$ gauge group and involves $N-1$ pairs of chiral
matter supermultiplets
 $S_{i\alpha}$, $S_i^{'\alpha}$
in the fundamental and anti-fundamental representations of the gauge
group with a common mass $m$.
 The lagrangian of the model reads
\beq
{\cal L} = \left( \frac{1}{4g^2} \mbox{Tr} \int d^2\theta \ W^2\ + \
{\rm H.c.}
\right)\ + \  \sum_{i=1}^{N-1} \left[ \frac{1}{4}\int d^2\theta d^2\bar\theta\
\bar S_{i}  e^V S_{i}  \right. \nonumber \\
+ \left.  \frac{1}{4}\int d^2\theta d^2\bar\theta\
 S'_{i}  e^{-V} \bar S'_{i}
 - \frac m2 \left(  \int d^2\theta\  S'_i S_{i}
+\mbox{H.c.}\right) \right]\ ,
\label{LSQCD}
\eeq
  color and Lorentz indices are suppressed. In this case,
 the gauge symmetry is broken
completely and the theory involves a discrete set of vacuum states.
The presence of $ N$ chirally asymmetric  states
has been known for a long time. They are best seen in the weak coupling
limit $m \ll \Lambda_{SQCD}$ where the chirally asymmetric states involve
large vacuum expectation values of squark fields $\langle s_i\rangle  \ \gg \Lambda_{SQCD}$
and the low energy dynamics of the model is described in terms of the colorless
composite fields ${\cal M}_{ij} = 2S'_i S_j$. The effective lagrangian
presents a Wess--Zumino model with the superpotential
 \beq
  \label{Higgs}
  {\cal W} = - \frac 23 \frac{\Lambda_{\rm SQCD}^{2N + 1}}
  {{\rm det} {\cal M}} - \frac{m}{2} {\rm Tr}\ {\cal M}
  \eeq
  The second term in Eq.(\ref{Higgs}) comes directly from the lagrangian
  (\ref{LSQCD}) and the first term is generated dynamically by instantons.
  Assuming ${\cal M}_{ij} = X^2 \delta_{ij}$ and solving the equation
  $\partial {\cal W}/\partial \chi = 0$ ($\chi$ is the scalar component of the
  superfield $X$), we find $N$ asymmetric vacua
  \beq
  \label{vacchi}
  \langle\chi\rangle_k = \left( \frac 43 \frac{\Lambda_{\rm SQCD}^{2N+1}} m
  \right)^{1/2N}
e^{\pi i k/N}
  \eeq
  (the vacua ``k'' and ``k + N''  have the same value of the
  moduli $\langle\chi^2\rangle_k$ and  are physically equivalent).
  These vacua are characterized by a finite gluino condensate
  \beq
  \label{cond}
  \langle{\rm Tr}\ \lambda^2\rangle_k \  = \ 8\pi^2  m\langle\chi^2\rangle_k
  \eeq

 It was noted recently \cite{Kovner} that on top of (\ref{vacchi})
  also a chirally symmetric vacuum with the
zero value of the condensate exists.  It cannot be detected
in the framework
of Eq.(\ref{Higgs}) which was derived {\it assuming} that the scalar v.e.v.
and the gluino condensate are nonzero and large, but is clearly
seen if writing down
the effective lagrangian due to Taylor, Veneziano, and
Yankielowicz (TVY) \cite{TVY} involving also the composite field
 \beq
 \label{Phi}
\Phi^3 = \frac 3{32\pi^2} {\rm Tr}\ W^2
 \eeq
   The corresponding superpotential reads
\footnote{The factor $2/3$ in Eq.(\ref{TVY}) and the corresponding factor in
Eq.(\ref{Higgs}) match the chosen (by historical reasons) normalization
factor in the definition (\ref{Phi}) of $\Phi$. }
  \beq
  \label{TVY}
{\cal W} =  \frac 23 \Phi^3 \left[ \ln \frac{\Phi^3 {\rm det} {\cal M}}
{\Lambda_{SQCD}^{2N + 1}} \ -\ 1 \right] - \frac{m}{2} {\rm Tr}\ {\cal M}
\eeq

The presence of different degenerate physical vacua in the theory
implies the existence of domain walls --- static field configurations
depending  only on one spatial coordinate ($z$) which interpolate between
one of the vacua at $ z = -\infty$ and another one at $z = \infty$ and
minimizing the energy functional. As was shown in \cite{Dvali}, in many
cases the energy density of these walls can be found exactly due to the
fact that the walls present the BPS--saturated states.

The energy density of a BPS--saturated wall in SQCD with $SU(N)$ gauge
group satisfies a relation \cite{my}
  \beq
   \label{eps}
   \epsilon \ =\ \frac {N}{8\pi^2} \left|\langle{\rm Tr}\ \lambda^2\rangle_\infty
    \ -\ \langle{\rm Tr}\ \lambda^2\rangle_{-\infty} \right|
    \eeq
where the subscript $\pm \infty$ marks the values of the gluino
condensate at spatial infinities.
\footnote{A relation of this kind can be derived also for other variants of
the theory involving  exotic groups and more complicated matter content, but
in general case the energy of a BPS wall and the gluino condensate are not
 related so directly \cite{G2}}.
 Bearing Eqs. (\ref{eps},\ \ref{cond}) in mind , the energy densities of the
BPS walls are
  \beq
 \label{epsr}
 \epsilon_r \ =\ N \left( \frac {4m^{N-1}}3  \right)^{1/N}
  \eeq
  for the  real walls and
  \beq
  \label{epsc}
\epsilon_c = 2 \epsilon_r  \sin \frac \pi N
 \eeq
 for the complex walls.
 The RHS of Eqs.(\ref{eps}-\ref{epsc}) presents an
absolute   lower bound for the energy of {\it any} field configuration
interpolating between different vacua.

    The relation (\ref{eps}) is valid {\it assuming} that the wall exists and
is BPS--saturated. However, whether such a BPS--saturated domain wall
exists or not is a non--trivial dynamic question which can be answered
only in a specific study of a particular theory in interest. This question
has already been studied in our previous works
\cite{my,SV,SVn,SUN}. In particular, in \cite{my,SV,SVn}  the simplest
model of the class (\ref{LSQCD}) with $N_c = 2,\ N_f = 1$ was analyzed.
The results are the following:
\begin{enumerate}
\item For any value of the mass of the matter
fields $m$, there are domain walls interpolating between a chirally asymmetric
and the chirally symmetric vacua (we call them {\it real} walls). They are
BPS -- saturated.
  \item There are also {\it complex} BPS solutions interpolating between
different chirally asymmetric vacua. But they exist only if the mass is small
 enough $m \leq m_* = 4.67059\ldots \Lambda_{SQCD}$. When $m > m_*$, BPS
  walls are absent.
  \item In a narrow range of masses $m_* < m \leq m_{**} \approx 4.83
\ \Lambda_{SQCD}$, complex domain walls still exist, but they are not BPS
saturated anymore. At $m > m_{**}$, there are no such walls whatsoever.
 \end{enumerate}
In Ref.\cite{SUN}, we studied the problem of existence of BPS--saturated
domain walls in the model (\ref{LSQCD}) with $N \geq 3$. The results are
basically the same as for $N=2$: the real walls exist for any $m$ and are
BPS--saturated, and there are two complex BPS branches which exist in a
limited range $m < m_*$. The value of $m_*$ goes down with $N$:
$ m_* = .28604\ldots \Lambda_{SQCD}$
for $SU(3)$, $ m_* = .07539\ldots \Lambda_{SQCD}$
for $SU(4)$, etc. For large $N$, $m_*(N) \propto N^{-3}$.

 The results concerning the BPS walls were obtained by solving numerically
 the first order BPS equations
   \beq
  \partial_z \phi \ =\ e^{i\delta} \partial \bar {\cal W} /\partial \bar
\phi,  \ \ \ \ \      \partial_z \chi \ =\ e^{i\delta}
\partial \bar {\cal W} /\partial \bar \chi\
  \label{BPS}
  \eeq
associated with the TVY lagrangian. The phase $\delta$ depends on  particular
vacua  between which the wall interpolates
(see Refs.\cite{my,Chib,SUN} for details).

To study the spectrum of the domain walls which are not BPS--saturated, one
has to solve the equations of motion which are of the second order, and,
technically, the problem is a little bit more involved. We did it earlier
for $N=2$ \cite{SVn}. This paper is devoted to the numerical solution of
the equations of motion for the $SU(3)$ gauge group.

\section{Solving equations of motion}

The scalar potential corresponding to the superpotential (\ref{TVY}) is
 \beq
U(\phi, \chi) \ =\ \left|\frac{\partial {\cal W}}{\partial \phi}
\right|^2 +
 \left|\frac{\partial {\cal W}}{\partial \chi}\right|^2 \ =\
4\left| \phi^2 \ln\{\phi^3 \chi^{2(N-1)}\} \right|^2 + (N-1)^2
\left|m\chi  - \frac{4\phi^3}{3\chi} \right|^2
\label{potTVY}
  \eeq
(from now on we set $\Lambda_{\rm SQCD} \equiv 1$).
  The potential (\ref{potTVY}) has $N+1$ degenerate minima. One of them is
  chirally
symmetric: $\phi = \chi = 0$.  There are also $N$ chirally asymmetric vacua
with $\langle\chi\rangle_k$ given in Eq.(\ref{vacchi}) and
   \beq
\langle\phi\rangle_k \ =\ \left( \frac {3m}4 \right)^{(N-1)/(3N)}
  e^{-\frac{2i(N-1)\pi k}{3N}}\
 \label{vacphi}
  \eeq

To study the domain wall configurations, we should add to the potential
(\ref{potTVY}) the kinetic term which we choose in the simplest possible
form
   \beq
\label{Lkin}
{\cal L}_{\rm kin} \ =\ |\partial \phi|^2
+ |\partial \chi|^2
  \eeq
  and solve the equations of motion with boundary conditions
  \beq
  \label{bc}
  \phi(-\infty) = \langle\phi\rangle_0 \equiv R_*, \ \ \phi(\infty) = R_*
  e^{-2\pi i (N-1)/3N}  \nonumber \\
  \chi(-\infty) = \langle\chi\rangle_0 \equiv \rho_*, \ \ \chi(\infty) = \rho_*
  e^{\pi i /N}
  \eeq
  Thereby we are studying the walls interpolating between ``adjacent'' complex
  vacua. Actual calculations will be performed for $N=3$ where all the
  vacua are adjacent. In principle, one could also study numerically the
  walls interpolating between the vacua $k=0$ and $k=2$ for, say, $SU(5)$
  gauge group, etc. We expect the physical results for all such walls to be
  qualitatively the same.

It is convenient to introduce the polar variables $\chi = \rho e^{i\alpha},\
\phi = R e^{i\beta}$ after which the equations of motion
acquire the form
 \beq
 \label{eqmot}
R'' - R \beta'^2 \ &=&\ 8R^3 [L(L + 3/2) + \beta_+^2] +
(N-1)^2 \left[\frac{16R^5}{3\rho^2} - 4mR^2 \cos (\beta_-) \right]
\nonumber \\
R\beta'' + 2R'\beta'\  &=& \ 12R^3\beta_+ + 4(N-1)^2 mR^2 \sin(\beta_-)
\nonumber \\
\rho'' - \rho \alpha'^2 \ &=&\ (N-1)\frac{8R^4}\rho L + (N-1)^2
\left(m^2\rho - \frac{16R^6}{9\rho^3} \right)  \nonumber \\
\rho \alpha'' + 2\rho' \alpha' \ &=&\ (N-1) \frac{8R^4}{\rho} \beta_+ -
(N-1)^2 \frac{8mR^3}{3\rho} \sin (\beta_-) \ ,
  \eeq
where $L = \ln[R^3 \rho^{2(N-1)}], \ \beta_+ = 3\beta + 2(N-1)\alpha,
\ \beta_- = 3\beta - 2\alpha$, with the boundary conditions
  \beq
  \label{bcpol}
\rho(-\infty)  =\rho(\infty) = \rho_*; \ \ R(-\infty) = R(\infty) = R_*;
\nonumber \\
\alpha(-\infty) = \beta(-\infty) = 0;\ \ \alpha(\infty) = \pi/N;\ \
\beta(\infty) = - \frac{2(N-1) \pi}{3N}
  \eeq
 When $N=2$, the system (\ref{eqmot}) is reduced to that studied in
Ref.\cite{SVn}.
 The system
(\ref{eqmot}) involves one integral of motion
   \beq
 \label{cons}
T - U \ =\ R'^2 + \rho'^2 + R^2 \beta'^2 + \rho^2 \alpha'^2 - \nonumber \\
 4R^4 (L^2 + \beta_+^2)
- (N-1)^2 \left[ m^2\rho^2 + \frac {16R^6}{9\rho^2} - \frac{8mR^3}3 \cos
(\beta_-) \right] \ =\ {\rm const}
  \eeq
  In our case, {\it const} = 0 due to boundary conditions (\ref{bcpol}).
The phase space of the system (\ref{eqmot}) is 8--dimensional
and a general Cauchy problem involves 8 initial conditions.
The  problem is simplified, however, when noting that the wall
 solution should be symmetric with respect to its center.
Let us seek for the solution centered at $z=0$ so that
\beq
\label{sym}
\rho(z)  = \rho(-z), \ R(z) = R(-z),\ \nonumber \\
\alpha(z)  = \pi/N - \alpha(-z),
\ \beta(z)  = -2(N-1)\pi/(3N) - \beta(-z)
\eeq
Indeed, one can be easily convinced that the Ansatz (\ref{sym}) goes through
the equations (\ref{eqmot}).
It is convenient to solve the equations (\ref{eqmot}) numerically on the
half--interval from $z=0$ to $z = \infty$. The symmetry (\ref{sym})
dictates $\rho'(0) = R'(0) =0,\ \alpha(0) = \pi/(2N),\ \beta(0) =
-(N-1)\pi/(3N)$ which fixes 4 initial conditions.
Four others satisfy the relation (\ref{cons}). Thus, we are left with 3 free
parameters, say, $\rho(0),\ R(0)$, and $\beta'(0)$, which should be fitted
so that the solution approach the complex vacuum in Eq.(\ref{bc}) at
$z \to \infty$.

 \begin{figure}
\grpicture{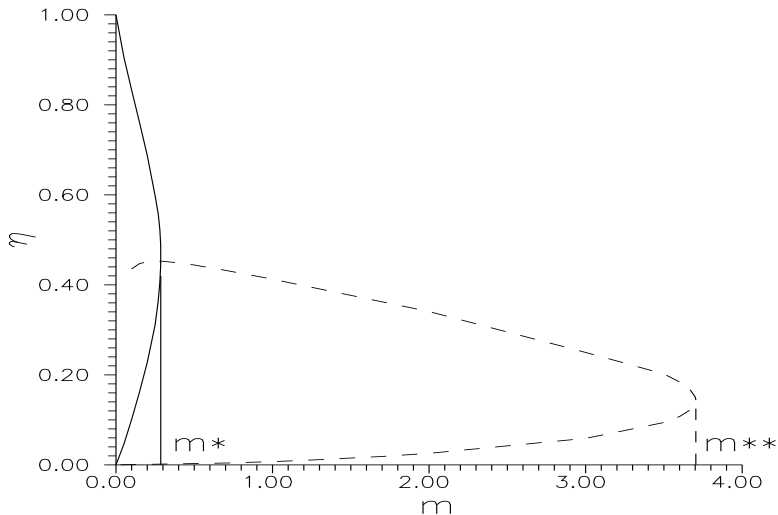}
\caption{The
ratio $\eta = R(0)/R_0(0)$  for the solutions of the equations of motion
as a function of mass for the
$SU(3)$ theory. The solid
lines describe the BPS solutions and the dashed lines describe the
non--BPS wall and the sphalerons. }
\label{vse}
\end{figure}

All the solutions obtained with such a procedure are presented in Fig.
\ref{vse} where the parameter $R(0)$, one of the fitted initial conditions,
is plotted as a function of mass (we normalized $R(0)$ at its value
$R_0(0) = .918\ldots R_* $ for the upper BPS branch in the limit
$m \to 0$;
see Ref. \cite{SUN} for details).  For small masses, there are several
solutions. We obtain first of all the solutions
studied in Ref.
\cite{SUN} and describing the BPS--saturated domain walls.
(solid lines in Fig. \ref{vse}). We find also two new
solution branches drawn with the dashed lines in Fig. \ref{vse}. We see that,
similarly to the BPS branches, two new dashed branches fuse together at some
$m = m_{**} \approx 3.704$ . No solution for the system (\ref{eqmot}) exist at
$m > m_{**}$.

 \begin{figure}
\grpicture{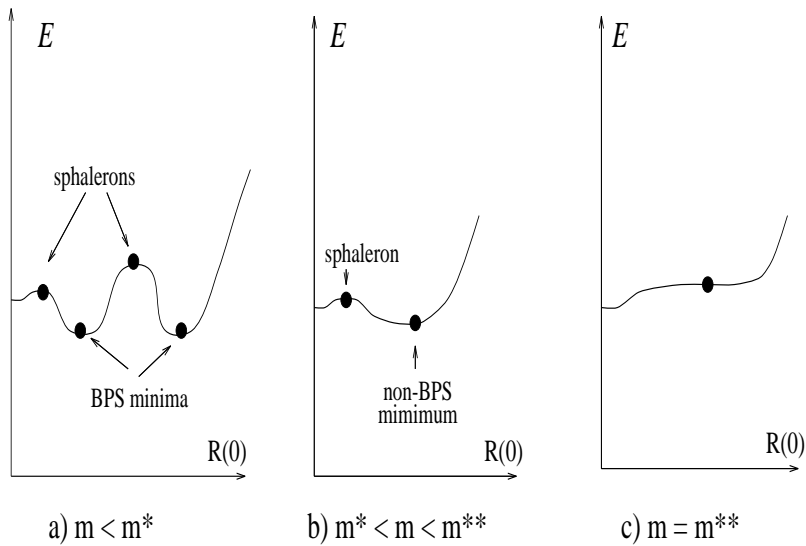}
\caption{Illustrative profiles of the energy functional vs. $R(0)$ .}
\label{Eplot}
\end{figure}

The picture is rather analogous to what we had for
$N=2$ \cite{SVn} and the physical
interpretation is similar. Let us assume first that
$m < m_*$ and draw the energy functional $ E$ for field configurations
with wall boundary conditions minimized over all parameters except the value
of $R(0)$  which is kept fixed (see Fig. \ref{Eplot}a). For very small
$R(0)$, our configuration
nearly passes the chirally symmetric minimum and the minimum
of the energy corresponds to two widely separated real walls. Thus
$ E[R(0) = 0] = 2\epsilon_r$ with $\epsilon_r$ given in Eq.(\ref{epsr}).
Two minima in Fig. \ref{Eplot}a correspond to BPS solutions with the
energy $\epsilon_c = \sqrt{3} \epsilon_r$. They are separated
by an energy barrier. The top of this barrier (actually, this is a saddle
point with only one unstable mode corresponding to $R(0)$, in other words
--- a {\it sphaleron})
is a solution described by the upper dashed line in Fig.\ref{vse}. The lower
dashed line corresponds to the local maximum on the energy barrier separating
the lower BPS branch and the configuration of two distant real walls at
$R(0) = 0$.
\footnote{In contrast to the case $N=2$, the existence of such a barrier
could not be established from the BPS spectrum alone. The matter is that,
while $2\epsilon_r = \epsilon_c$ for $N=2$ and the presence of the maximum
is guaranteed by the Roll theorem, $2\epsilon_r > \epsilon_c$
in our case and one could in principle imagine a situation where
$ E$ falls down monotonically when $R(0)$ is increased from zero
up to its value at the lower BPS branch. As will be discussed later, our
numerical results are not good enough to make a {\it rigid} statement on the
form of the function $ E[R(0)]$ at small $R(0)$ in the region $m < m_*$,
but they strongly suggest that the energy barrier (
though a tiny one) is present. }

 At $m = m_*$,
two BPS minima fuse together and the energy barrier separating them
disappears. The upper sphaleron branch coincides
with the BPS solution at this point. When $m$ is increased above $m_*$, the
former BPS minimum  is still a  minimum of the energy functional, but its
 energy is now above the BPS bound (see Fig.\ref{Eplot}b).
The corresponding solution is described by the analytic continuation of the
upper sphaleron branch. The lower dashed branch in the region
$m_* < m < m_{**}$ is still a sphaleron. At the second critical point
$m = m_{**}$, the picture is changed again (see Fig.\ref{Eplot}c). The local
maximum and the local minimum fuse together and
the only one remaining stationary point does not correspond to an extremum
of the energy functional anymore. At larger masses, no non-trivial stationary
 points are left.

 \begin{figure}
\grpicture{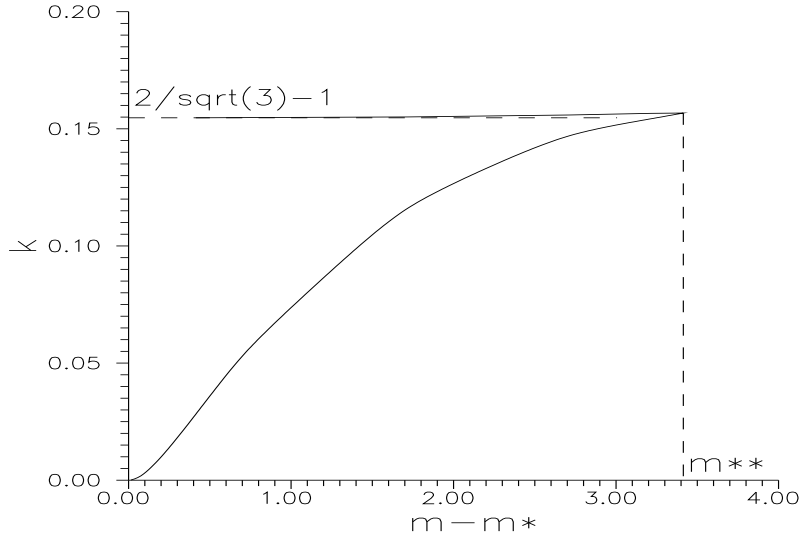}
\caption{The ratio $\kappa(m) = E/\epsilon_c -1$ for the non--BPS wall
and for the lower sphaleron as a function of mass.}
\label{Edve}
\end{figure}

 \begin{figure}
\grpicture{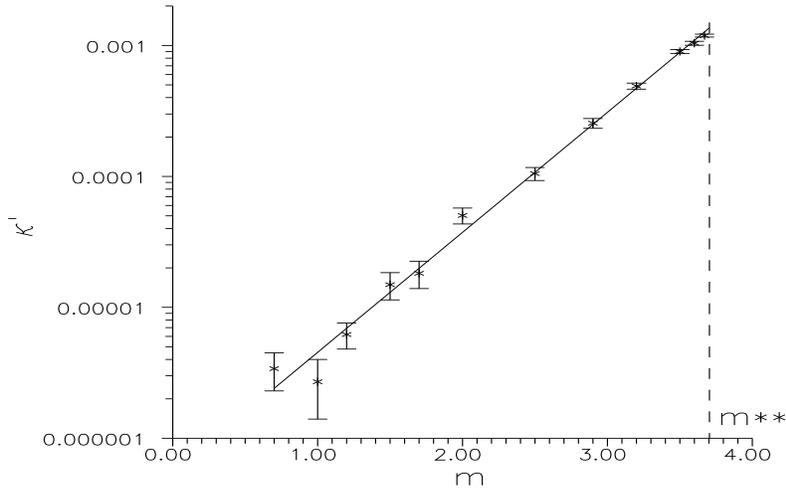}
\caption{Logarithmic plot for the ratio $\kappa'(m) =
E_{\rm lower\ sphaleron}/(2\epsilon_r) - 1$
 vs. mass.}
\label{Elog}
\end{figure}

Our findings are illustrated in Figs. \ref{Edve}, \ref{Elog} where the
energies of the non-BPS wall and of the lower sphaleron branch are plotted
as a function of $m$. In Fig. \ref{Edve}, the ratios of the energies of both
branches to the BPS bound (\ref{epsc}) are plotted. The lower line in Fig.
\ref{Edve} corresponds to the  stable wall solution and the upper
line to the sphaleron branch. For almost all $m < m_{**}$, the wall solution
is globally stable. When $m_{**} - m$ becomes very small, it is stable
only locally: we see that,
at $m = m_{**}$ where two branches are fused together, their energy exceeds
slightly the energy of two real walls $2\epsilon_r$.

In Fig. \ref{Elog}
the sphaleron energy is redrawn in logarithmic scale in the
units of $2\epsilon_r$.
Unfortunately, we do not have good numerical data at $m \lsim .7$ --- our
relative uncertainty becomes large. We see, however, that the logarithmic plot
in Fig. \ref{Elog} is pretty much linear, the best fit is
 \beq
 \label{fit}
 \kappa'(m) = 5.49 \cdot 10^{-7}  \exp\{2.11 m \}
 \eeq
  The fit (\ref{fit}) cannot be valid for very small masses: we
  expect $\kappa'(0) = 0$ which means that the straight line in Fig.
  \ref{Elog} should bend down at small enough masses due to a preexponential
  factor $\sim m^\alpha$ which we cannot determine from our data. Anyway,
   it is seen
  from Fig. \ref{Elog} that $\ln \kappa'$ does not ``want'' to hit infinity
  at a finite mass, and we assume that it does not (
  though we cannot exclude it as a logical possibility).
  In terms of Fig. \ref{Eplot}, that means that
   the  energy barrier on the left
(for illustrative purposes, it is very much exageratted ) is present
for all masses.

 \begin{figure}
\grpicture{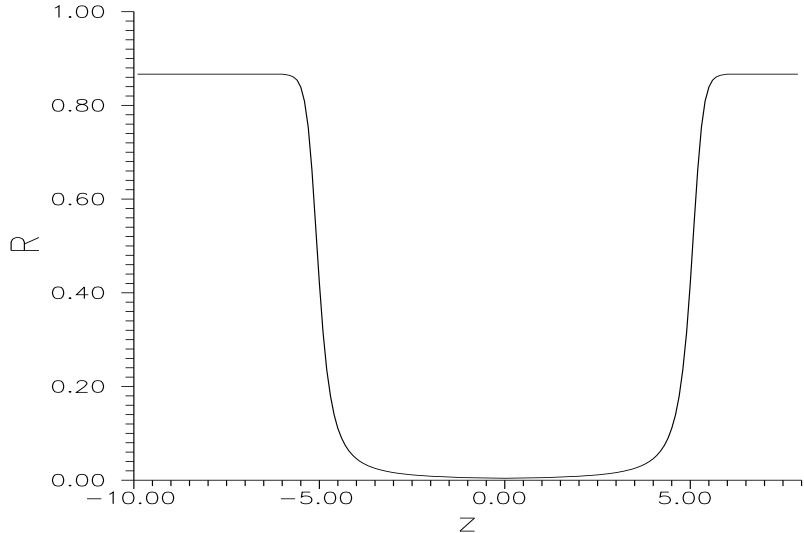}
\caption{Profile $R(z)$ for the lower sphaleron branch at $m = .7$ .}
\label{prof}
\end{figure}

Finally, we present for illustration the profile $R(z)$ for
the lower sphaleron branch at $m = .7$. As was expected, it resembles very
much a combination of two separate real domain walls.

\section{Discussion.}

Our main conclusion is that, besides the critical mass $m_*$ beyond which
BPS solutions disappear, also a second critical mass $m_{**}$ exists beyond
which no complex wall solution can be found whatsoever. This was the case
for $SU(2)$ \cite{SVn} and, as we see now, this is also the case for $SU(3)$.
Seemingly, the same situation holds for any $N$.
That means in particular that no domain walls connecting different chiraly
asymmetric vacua are left in the pure supersymmetric Yang--Mills theory
corresponding to the limit $m \to \infty$, and only the real domain walls
connecting the chirally symmetric and a chirally asymmetric vacuum states
survive in this limit. That contradicts an {\it assumption} of
Ref.\cite{Witten} that it is complex rather than real
domain walls which are present in the pure SYM theory (Witten discussed them
in the context of brane dynamics).

One has to
make a reservation here: our result was obtained in the framework of the
TVY effective lagrangian (\ref{TVY}) whose status [in contrast to that
 of the lagrangian (\ref{Higgs})] is not absolutely clear:
the field $\Phi$ describes heavy  degrees of freedom (viz. a scalar
glueball and its superparnter) which are not nicely separated
from all the rest. However, the form of the superpotential (\ref{TVY})
and hence the form of the lagrangian for static field configurations
is {\it rigidly} dictated by symmetry considerations; the uncertainty
involves only kinetic terms. It is reasonable to assume that, as far as
the vacuum structure of the theory is concerned (but not e.g. the excitation
spectrum --- see Ref.\cite{SVn} for detailed discussion), the effective TVY
potential (\ref{TVY}) can be trusted. A recent argument against using
Eq. (\ref{TVY}) that the chirally symmetric phase whose existence follows
from the TVY lagrangian does not fulfill certain discrete anomaly matching
conditions \cite{Csaki} is probably not sufficient. First, it assumes that
the excitation
spectrum in the symmetric phase is the same as it appears in the TVY
lagrangian which is not justified. Second, it was argued  recently
that the TVY lagrangian describes actually {\it all} the relevant symmetries
of the underlying theory and the absence of the anomaly matching is in a sense
an ``optical illusion'' \cite{KKS}.

The main distinction of the $SU(3)$ case considered here compared to the
$SU(2)$ theory is that the values of two critical values are rather
different (in $SU(2)$ case they were pretty close: $m_* = 4.67059\ldots$
and $m_{**} \approx 4.83$). This is due to the fact that the energy
of the complex BPS wall $\epsilon_c$ is less in this case than the energy
of two real walls $2\epsilon_r$. When we increase the mass and go
above $m_*$, the energy of the wall first has to rise from $\epsilon_c$
to $2\epsilon_r$. Only then the  complex domain
wall ``bound state'' can break apart into its ``constituents'', the real walls.
\footnote{Actually, as we have seen, the energy barrier in Fig. \ref{Eplot}
does not allow the complex wall to break apart until its energy goes
a little bit above the limit $2\epsilon_r$.}

One can expect that $m_*$ and $m_{**}$ differ more and more as $N$ grows.
A tentative guess is that $m_{**}$ is roughly $N$--independent
(to be compared with $m_*(N) \propto N^{-3}$). Of course, that can be confirmed
or disproved by only actual numerical study. Note, however, that numerical
calculations become more and more difficult as $N$ grows --- the instabilities
characterized by eigenvalues of the Jacobi matrix of the system (\ref{eqmot})
near the minima (\ref{vacchi}, \ref{vacphi}) grow as $N^2$.

\vspace{.5cm}

{\bf Acknowledgments}: \hspace{0.2cm}
This work was supported in part  by the RFBR--INTAS grants 93--0283, 94--2851,
95--0681, and 96-370, by the RFFI grants 96--02--17230,  97--02--17491, and
97--02--16131, by the RFBR--DRF grant 96--02--00088,
by the U.S. Civilian Research and Development Foundation under award
\# RP2--132, and by the Schweizerishcher National
Fonds grant \# 7SUPJ048716.

\end{document}